\documentstyle[aps,pra,epsf]{revtex}
\begin{document}
\draft
\newcommand{\pp}[1]{\phantom{#1}}
\newcommand{\be}{\begin{eqnarray}}
\newcommand{\ee}{\end{eqnarray}}
\newcommand{\ve}{\varepsilon}
\newcommand{\vs}{\varsigma}
\newcommand{\Tr}{{\,\rm Tr\,}}
\newcommand{\pol}{\frac{1}{2}}
\newcommand{\ba}{\begin{array}}
\newcommand{\ea}{\end{array}}
\newcommand{\bear}{\begin{eqnarray}}
\newcommand{\eear}{\end{eqnarray}}

\title{Relativistic corrections to the Ekert test for eavesdropping}
\author{Marek Czachor}
\address{
Katedra Fizyki Teoretycznej i Metod Matematycznych\\
 Politechnika Gda\'{n}ska,
ul. Narutowicza 11/12, 80-952 Gda\'{n}sk, Poland}

\maketitle
\begin{abstract}
 A degree of violation of the Bell
inequality depends on momenta of massive particles with
respect to a laboratory if spin plays a role af a  ``yes--no" 
observable. For
ultra-relativistic particles a standard Ekert test has to take
into accont this velocity dependent suppression of the degree of
violation of the inequality. Otherwise Alice and Bob may
``discover" a nonexisting eavesdropper. 
\end{abstract}

\bigskip
The cryptographic protocol proposed by Ekert \cite{[1]}
consists of two essential parts. First, a two-particle
singlet-state source produces pairs of particles generating
identical sequences of bits which can be subsequently 
used by Alice and Bob
as a cryptographic key. Second, some randomly chosen data are
used to test whether the particles were really prepared in a
singlet state. This can be done by measuring the ``Bell average"
$$
c(a,a',b,b',\psi)=\langle\psi|\hat a\otimes \hat b
+
\hat a\otimes \hat b'
+
\hat a'\otimes \hat b
-
\hat a'\otimes \hat b'
|\psi\rangle
$$
where the binary observables $\hat a$, etc., are chosen in such a way
that the Bell inequality $|c(a,a',b,b',\psi)|\leq 2$ is violated.
This will be done most efficiently  if $|c(a,a',b,b',\psi)|=
2\sqrt{2}$ and the violation is expected to be maximal. A
result of the form $|c(a,a',b,b',\psi)|< 2\sqrt{2}$ means that
$|\psi\rangle$ is not a pure singlet, and this indicates 
an eavesdropper. After having detected the eavesdropper the data
collected by Alice and Bob are thrown away and the whole
procedure starts again. 
\bigskip
The binary observables appearing in $c(\dots)$ represent results of some
``yes--no" experiments and it is known that for any
entangled state there exists a set of observables leading to the
maximal violation of the inequality. This statement is, strictly
speaking, true {\it in principle\/}. In practice, however, there
is no operational rule that could tell us how to measure an
arbitrary observable. So practically we are  confined to a
concrete set of ``physical" observables which, in this context,
are various ``polarizations". In the photon case the
``polarizations" are either simply linear polarizations or some
observables related to interferometric experiments. The
situation is less clear if one considers massive particles. 
\bigskip
At this moment one may ask why should one worry about massive particles if
massless photons work so well. A practical answer is provided
by a very high efficiency of atomic detectors as compared to
the photonic ones \cite{[2]}. 
Taking into account the spectacular comeback of
atoms in interfereometry  and the recent 
progress in experiments with coherent atomic sources 
one cannot exclude future applications of atomic entangled
states in cryptography. The question I want to discuss here is
the role of relativistic effects that may become important
if very fast atomic EPR-type beams are used for
the cryptographic key distribution \cite{[3]}. 
\bigskip
Let us begin with the yes--no polarizations for relativistic
massive particles. It is known that the spinor part $\vec s$ of the
generator of rotations is not a well defined spin operator for,
say, the Dirac electrons. Its projection on the momentum
direction (helicity) commutes with the free Dirac Hamiltonian,
but to measure $c(\dots)$ we need measurements of spin in
different directions. The naive approach based on $\hat a=\vec
a\cdot\vec s$ cannot be physically meaningful since eigenstates
of $\hat a$ are not preserved by a free evolution.
This is a general phenomenon present in all Poincar\'e covariant
systems but this argument alone does not prove that any notion
of spin is meaningless. There are at least two objects whose
definitions are 
representation-independent and which  can represent a physical
spin of a relativistic system. One is the spin operator $\vec S$
constructed by means of a relativistic center-of-mass position
operator. The other is the Pauli-Lubanski vector $(W_0,\vec W)$.
The two objects are not equivalent but their spacelike parts 
are proportional to each another. 
Their projections on $\vec a$ have different eigenvalues but the
same eigenstates. Accordingly they both give the same values of
$c(\dots)$. 
\bigskip
Let $a_\alpha=(a_0,\vec a)$ be some world-vector. One can
consider joint eigenstates of the Pauli-Lubanski vector and the
4-momentum. The eigenvalues of
$a_\alpha W^\alpha=a\cdot W$ are 
$$
w_\pm(a,p)=\pm j_3\sqrt{(a\cdot p)^2 - a^2p^2}=\pm w(a,p)
$$
where $j_3=1/2,1,3/2,\dots,j$, and $a^2=a_\alpha
a^\alpha$ etc. For $j=1/2$ the ``yes--no" observable $\hat a$ can be defined
in a relativistically invariant way as 
$$
\hat a=\hat a(p)={a\cdot W\over w(a,p)}.
$$
An arbitrary two-spin-1/2-particle state can be written as 
$$
|f\rangle =
\sum_{\sigma_1,\sigma_2}\int d\Gamma(\vec p_1)d\Gamma(\vec p_2)\,
f(\sigma_1,\vec p_1,\sigma_2,\vec p_2)|\sigma_1,\vec p_1\rangle
|\sigma_2,\vec p_2\rangle 
$$
where $\sigma_k$ is a helicity,
$f(\sigma_1,\vec p_1,\sigma_2,\vec p_2)=-f(\sigma_2,\vec
p_2,\sigma_1,\vec p_1)$, 
and $d\Gamma(\vec p)=(2\pi)^{-3}(2|p_0|)^{-1}d^3p$ is an
invariant measure on a mass hyperboloid. 
Let us consider the particular state 
$$
|f\rangle =
{1\over\sqrt{2}}
\int d\Gamma(\vec p_1)d\Gamma(\vec p_2)\,
f(\vec p_1,\vec p_2)\Bigl(|+,\vec p_1\rangle
|-,\vec p_2\rangle
-
|-,\vec p_1\rangle
|+,\vec p_2\rangle \Bigr)\eqno{(1)}
$$
where $f(\vec p_1,\vec p_2)=f(\vec p_2,\vec p_1)$. Denote
$\vec \beta=\vec p/p_0$, $\vec n=\vec p/|\vec p|$, and consider
projections of $W^\alpha$ on spacelike directions satisfying in
a laboratory frame $a_\alpha=(0,\vec a)$, $b_\alpha=(0,\vec b)$.
The laboratory frame is here the one which defines the
decomposition of the 4-momentum into energy and 3-momentum $\vec
p$ appearing in the wave functions. 
The EPR average is 
\be
\langle f|\hat a\otimes \hat b|f\rangle 
&=&
-
\int d\Gamma(\vec p_1)d\Gamma(\vec p_2)\,
|f(\vec p_1,\vec p_2)|^2\nonumber\\
&\pp=&\times
\frac{
\Bigl(
\sqrt{1-\beta^2_1}\,\bigl[\vec a-(\vec a\cdot \vec n_1)\vec n_1\bigr] + 
\bigl(\vec a\cdot \vec n_1\bigr)\vec n_1\Bigr)\cdot
\Bigl(
\sqrt{1-\beta^2_2}\,\bigl[\vec b-(\vec b\cdot \vec n_2)\vec n_2\bigr] + 
\bigl(\vec b\cdot \vec n_2\bigr)\vec n_2\Bigr)}{
\sqrt{1+\beta_1^2\bigl[(\vec n_1\cdot \vec a)^2-1\bigr]}\,
\sqrt{1+\beta_2^2\bigl[(\vec n_2\cdot \vec b)^2-1\bigr]}}
\ee
For 
$$
d\Gamma(\vec p_1)d\Gamma(\vec p_2)\,
|f(\vec p_1,\vec p_2)|^2
\approx
d^3p_1d^3p_2 \delta(\vec p-\vec p_1)
\delta(\vec p_1-\vec p_2)\eqno{(2)}
$$
we obtain the result discussed in \cite{[3]}:
$$
\langle f|\hat a\otimes \hat b|f\rangle 
\approx
-
{
\vec a\cdot \vec b - \beta^2 \vec a_\perp\cdot \vec b_\perp\over 
\sqrt{1+\beta^2\bigl[(\vec n\cdot \vec a)^2-1\bigr]}\,
\sqrt{1+\beta^2\bigl[(\vec n\cdot \vec b)^2-1\bigr]}}.\eqno{(3)}
$$
(3) implies 
$$
\langle f|\hat a\otimes \hat a|f\rangle =-1,
$$
which means that spins are always anti-parallel. The state
(1) satisfying (2)  
can be therefore used for a secure key transfer. The same is
true for more realistic states satisfying 
$$
d\Gamma(\vec p_1)d\Gamma(\vec p_2)\,
|f(\vec p_1,\vec p_2)|^2
\approx
d^3p_1d^3p_2 |f(\vec p_1)|^2
\delta(\vec p_1-\vec p_2)
$$
where $|f(\vec p_1)|^2$ is a probability density. Notice that
for $\vec a=\vec a_\perp$, $\vec b=\vec b_\perp$, we obtain the
nonrelativistic formula 
$$
\langle f|\hat a\otimes \hat b|f\rangle =-\vec a\cdot\vec b.
$$
The condition $\vec a=\vec a_\perp$, etc., cannot be satisfied
in a real experiment since $\vec a$ is a fixed vector and the
set of momenta perpendicular to a given fixed axis has
3-dimensional measure 0. So from any realistic wave packet one
can remove all $\vec p$ perpendicular to $\vec a$ and an
experimentally measured average will not change. So assume that
$\vec a\cdot \vec p\neq 0$. In this case we find that 
$$
\langle f|\hat a\otimes \hat b|f\rangle \to-1
$$
for $\beta\to 1$ independently of $\vec a$, $\vec b$. Here
$\beta$ is the average velocity of the wave packet. 
\bigskip
To better understand this result let us recall the explicit form
of the relativistic spin operator 
$$
\vec S=\vec W/p_0=\sqrt{1-\beta^2}\,\vec s_\perp +(\vec n\cdot
\vec s)\vec n.
$$
In the $\beta \to 1$ limit the vector points in the momentum
direction and 
$$
{[S_k,S_l]}\to 0.
$$
The relativistic spin becomes the more ``classical" the faster
the particle. This explains qualitatively why one should expect a
{\it suppression of the Bell inequality violation\/} for
ultra-relativistic massive particles [3]. It has been
shown in [3] that for $\beta \to 1$ 
$$
|c(a,a',b,b',f)|\to 2
$$
from above even if the measuring devices are optimally 
chosen. 

The plots illustrate these phenomena. The vectors in
$c(\dots)$ are  chosen coplanarly in a way which leads to the maximal
violation of the Bell inequality in the nonrelativistic case
($|c|=2\sqrt{2}$): $\vec a=(1/\sqrt{2},1/\sqrt{2},0)$, 
$\vec a'=(-1/\sqrt{2},1/\sqrt{2},0)$, $\vec b=(0,1,0)$, and 
$\vec b'=(1,0,0)$. The measuring devices are at rest in a
laboratory frame and $\vec \beta$ denotes velocities of
particles with respect to the lab. 

One may wonder why do we have here a suppression of
$|c(a,a',b,b',f)|$ for any spin and simultaneously have maximal
violation for photons. The answer is that in the photon
polarization or interferometric experiments we do not measure
spins. So a possibility exists that there are also some other
observables that could be used in the EPR-type experiments with
massive particles instead of spin, but none have been proposed
so far. 

The moral following from our story is that Alice and Bob do not
necessarily have to throw away all data if they find
$|c(a,a',b,b',f)|<2\sqrt{2}$. They first have to check the
relativistic corrections. In order to do this they have to know
the momentum distribution of the particle beam. The momentum can
be measured, in principle, on each particle without disturbing
its spin since 4-momentum and $W_\alpha$ commute. 

\acknowledgments
This work was partially supported by the KBN grant 2P30B03809
and is a part of the Polish-Flamish project on ``New models of
probability calculus and new experiments on single quantum
particles".  I'm indebted to the Batory Foundation and
Nokia--Poland for financial support. 
\begin{figure}
\epsffile{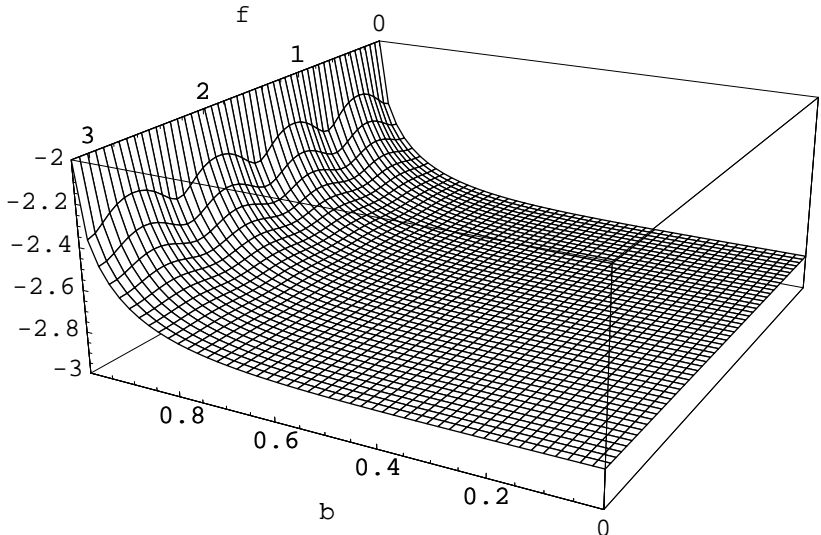}
\caption{
Both particles move with velocity $\vec \beta=
\beta(\cos\phi,\sin\phi,0)$. 
The maximal violation
is found for $\beta=0$ and there is no violation
for $\beta\to 1$. }
\end{figure}
\begin{figure}
\epsffile{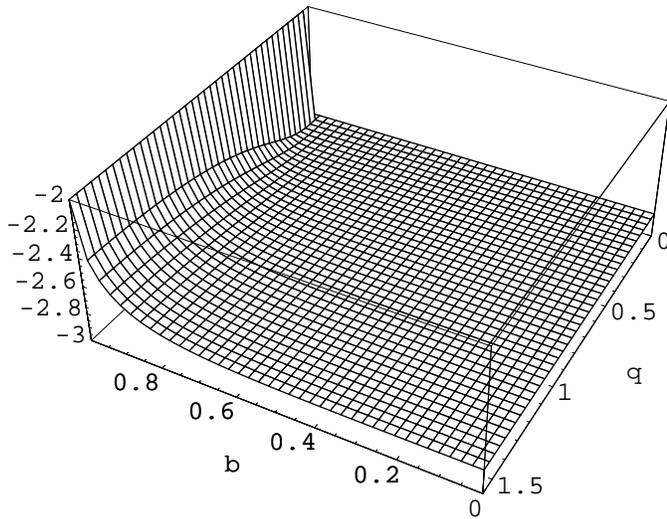}
\caption{
Similar situation, but now the
particles move with velocity $\vec \beta=
\beta(\sin\theta,0,\cos\theta)$. 
The maximal violation is found
for $\beta=0$ and is suppressed as $\beta$ approaches 1. For $\theta=0$
the particles move perpendicularly to the measuring devices and
the violation remains maximal for all $\beta$. The set of all such
momenta is of measure 0 which implies that one should
expect the relativistic damping effect to be present for all
wave packets.}
\end{figure}
\begin{figure}
\epsffile{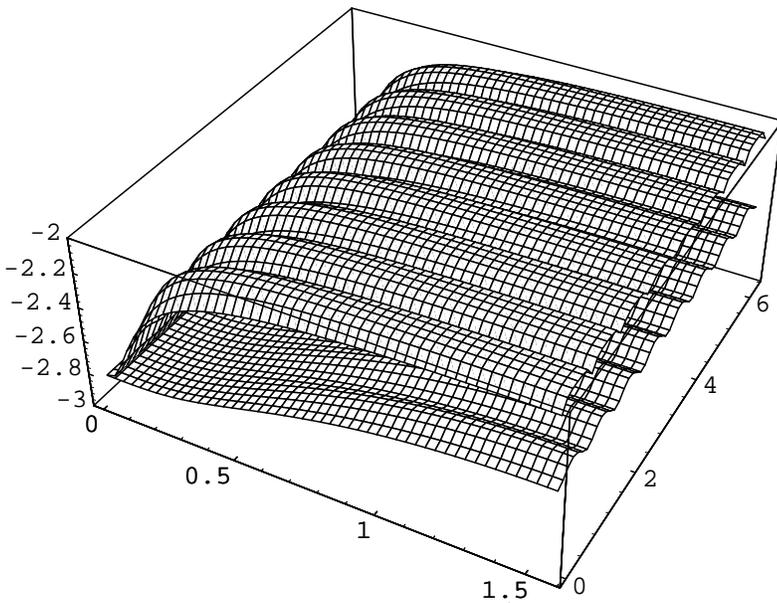}
\caption{
Situation from Fig.~1 but with $\vec \beta=
\beta(\cos\phi\sin\theta,\sin\phi\sin\theta,\cos\theta)$ for
$\beta=0.99$ (upper) and $\beta=0.95$ (lower); $0\leq\phi\leq
2\pi$, $0\leq\theta\leq \pi$. 
}
\end{figure}
\begin{figure}
\epsffile{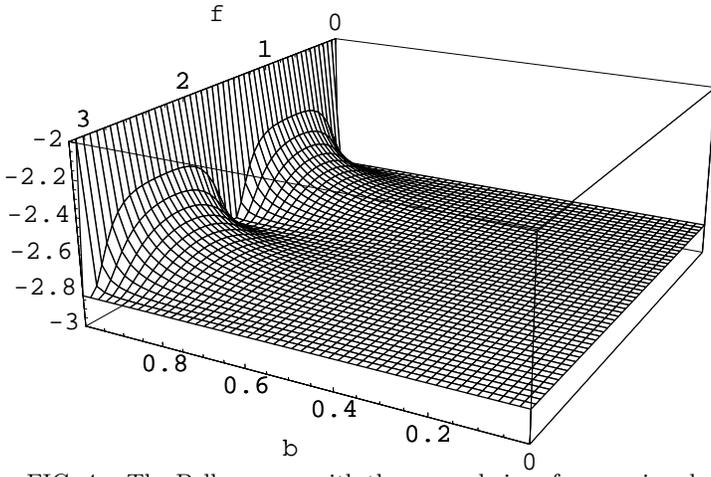}
\caption{
The Bell
average with the same choice of measuring devices but now the 
velocities of the two particles are different: One moves
nonrelativistically ($\beta_1\approx 0$) and the other with an
arbitrary velocity $\vec \beta_2=\beta(\cos\phi,\sin\phi,0)$. 
The plot is very similar to the one from Fig.~1. }
\end{figure}
\begin{figure}
\epsffile{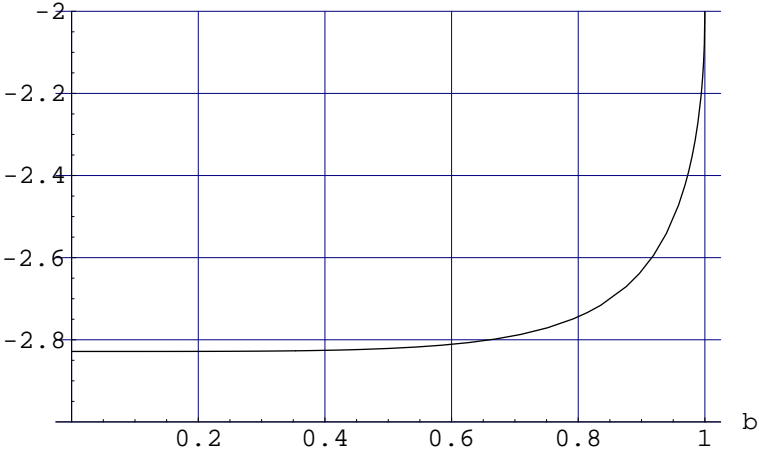}
\caption{
A cut through
the surface from Fig.~1 at the ``gully" $\phi=0$. }
\end{figure}
\begin{figure}
\epsffile{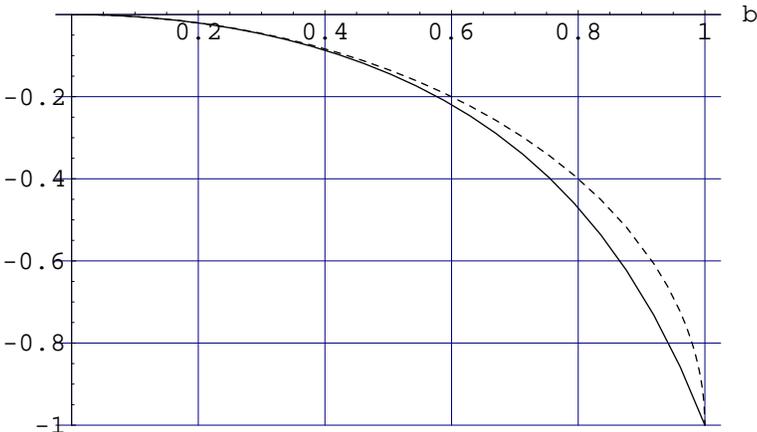}
\caption{
The relativistic corrections are much more visible if
instead of measuring the entire Bell average one concentrates 
on $\langle f|\hat a\otimes \hat b|f\rangle$ and takes $\vec a$
and $\vec b$ perpendicular. For particles moving
with velocity $\vec \beta=\beta\vec n$, $\vec a\cdot\vec b=0$, 
$\vec a\cdot\vec n=\vec b\cdot\vec n=2^{-1/2}$ we find
$\langle f|\hat a\otimes \hat
b|f\rangle=-\beta^2(2-\beta^2)^{-1}$. The solid curve shows this
average as compared to $[1-\beta^2]^{1/2}-1$ (dotted). }
\end{figure}

\end{document}